\documentstyle[prd,aps,epsf]{revtex}
\begin{document}
\draft


\title{Classical Behaviour After a Phase Transition:\\
             II. The Formation of Classical Defects}

\author{R.\ J.\
Rivers $^{1}$\thanks{r.rivers@ic.ac.uk}, F. \ C.\ Lombardo$^{1,2}$
\thanks{f.lombardo@ic.ac.uk}, and F. \ D.\
Mazzitelli$^{2}$\thanks{fmazzi@df.uba.ar}}

\address{{\it
$^1$ Theoretical Physics Group; Blackett Laboratory, Imperial
College,
London SW7 2BZ\\
$^2$ Departamento de F\'\i sica, Facultad de Ciencias Exactas y Naturales\\
Universidad de Buenos Aires - Ciudad Universitaria,
Pabell\' on I\\
1428 Buenos Aires, Argentina}}

\maketitle

\begin{abstract}

Classical defects (monopoles, vortices, etc.) are a characteristic
consequence of many phase transitions of quantum fields. Most
likely these include transitions in the early universe and such
defects would be expected to be present in the universe today. We
continue our analysis of the onset of classical behaviour after a
second-order phase transition in quantum field theory and show how
defects appear after such  transitions.

\end{abstract}


\vskip2pc

\section{Introduction}

Because phase transitions take place in a finite time causality
guarantees that, even for continuous transitions, correlation
lengths remain finite. Order parameter fields become frustrated,
and topological defects arise so as to reconcile field phases
between different correlated regions \cite{kibble1,zurek1}.

A huge variety of defects is possible, according to the complexity
of the initial symmetry and  its breaking \cite{kibble2}. Monopoles can
easily overwhelm the energy density of the universe, while cosmic strings
(vortices) may be the source of high energy cosmic rays and
lensing, as well as contributing to structure formation (although
they are not now thought to be its determining factor). More
complex strings can exacerbate baryogenesis.

As solutions to the classical field equations, these defects have
non-perturbatively large energies, commensurate with the
temperature scale at which the transition takes place. Thus, for
example, cosmic strings produced in the early universe will be
expected to have an energy/unit length (tension) $\sigma\sim
(10^{16}GeV)^2$, where $10^{16}GeV$ is the estimated GUT scale.
For this reason alone, defects are manifestly classical entities
after the transition, whose evolution can be handled successfully
through the solution of the classical equations that they satisfy
(e.g. the Nambu-Goto action for cosmic strings).  Nonetheless,
their origin in the beginnings of the phase transition that
spawned them is quantum mechanical. There are further
complications according as the symmetries are global or gauged
\cite{rajantie}, but the simplest of all are topological global
vortices (or strings) and it is these that we shall consider here.

Having set up the model, the subsequent sections of this work
concern the transition from quantum field theory to classical
field theory. We know the mechanism for this, the decohering
effects of the quantum environment with which the long-wavelength
modes of the order parameter fields interact \cite{fernando2}.
Field or phase ordering and classical defect formation are
controlled by long-wavelength modes. We consider these modes to
constitute the open 'system' which undergoes quantum decoherence
due to an 'environment' which consists of everything else
(short-wavelength modes and all other fields interacting with the
order parameter fields). This permits the calculation of a
decoherence time $t_D$ after the onset of the transition, after
which the long-wavelength modes behave classically, subject to
generalised Langevin equations. At the same time, the Wigner
functional can now be interpreted as a Fokker-Planck probability
density and we can calculate correlation functions \cite{diana}.
Given that the correlation functions are now defined in terms of
classical probabilities, classical defects can be identified.

The major characteristic of defects is their topological charge.
We conclude by showing how localised topological charge precedes
the appearance of defects, and gives an estimate of defect
densities when they do appear.

For an instantaneous quench the onset of classical behaviour has
already been discussed by us in an earlier paper in these
proceedings, henceforth known as I (see \cite{fernando1}), and
elsewhere\cite{lmr}, and we shall not repeat all the details here.
What we shall do is elucidate those properties of the
classicalisation due to the environment that are relevant to
defect formation.

We conclude by briefly comparing our approach to others for the
production of defects.  This work has yet to be completed, and
will be continued elsewhere.

\section{The model: Coupling order-parameter fields to the environment}

The most immediate difference between the model here and that of
paper I \cite{fernando1} lies in the replacement of the single
real scalar order parameter field there by a complex field $\Phi$.
This is the simplest field permitting acceptable defects, in the
form of global vortices. [The defects of a single scalar field
with broken $Z_2$ symmetry, domain walls, would dominate the
energy density of the universe if they were present.] As in I we
take the $\Phi$ field to interact with a collection of real scalar
environmental fields $\chi_{\rm n}$ (n=1,...,N). It is no loss of
generality to take all couplings to $\chi$-fields identical, and
all $\chi_{\rm n}$ masses identical, to give a model with
$U(1)\times O(N)$ or $O(2)\times O(N)$ symmetry broken to $O(N)$.
However, the $O(N)$ symmetry of the model, while tactically
useful, is accidental and purely for calculational convenience. We
adopt a Cartesian field basis $\Phi = (\phi_1 +i\phi_2)/\sqrt{2}$
because, in the linear regime at the early times that are most
relevant to us, $\phi_1$ and $\phi_2$ behave independently. The
action is
\begin{equation}
S[\Phi , \chi ] = S_{\rm syst}[\Phi ] + S_{\rm env}[\chi ] +
S_{\rm int}[\Phi ,\chi ], \label{action1}
\end{equation}
where we have made a provisional separation into system field and
environment with which it interacts, where
\[
S_{\rm syst}[\Phi ] = \int d^4x\left\{ {1\over{2}}\partial_{\mu}
\phi_{\rm a}\partial^{\mu} \phi_{\rm a} + {1\over{2}}\mu^2
\phi_{\rm a}^2 - {\lambda\over{4}}(\phi_{\rm a}^2)^2\right\}
\]
is
the action for the $\phi_{\rm a}$ fields (${\rm a}=1,2$) (we
assume summation over repeated index ${\rm a}$), with $\mu^2
>0$, and
\begin{eqnarray} S_{\rm env}[\chi ]
&=& \sum_{\rm n=1}^N\int d^4x\left\{
{1\over{2}}\partial_{\mu}\chi_{\rm n}
\partial^{\mu}
\chi_{\rm n} - {1\over{2}} m^2 \chi^2_{\rm n}\right\}, \nonumber
\\
S_{\rm int}[\Phi ,\chi ] &=& - \sum_{\rm n=1}^N\frac{g}{8\sqrt{N}}
\int d^4x \,\phi_{\rm a}^2 (x) \chi^2_{\rm n} (x), \nonumber
\end{eqnarray}
describe the environmental fields $\chi_{\rm n}$ and its
interaction with them. These are taken to be weakly coupled, with
$1\gg 1/\sqrt{N}\gg g\simeq\lambda$. Meanwhile, for simplicity the
$\chi_{\rm n}$ masses are fixed at $ m\simeq\mu$.

The model has a continuous transition for the breaking of the
$O(2)$ symmetry at the critical temperature $T_{\rm c}$ where, in
units in which $k_B = 1$, $T_{\rm c}^2 = {\cal
O}(\mu^2/\sqrt{N}\lambda)\gg\mu^2$. We shall assume that the
initial states of the system and environment are both thermal, at
temperatures $T=O(T_{\rm c})>T_{\rm c}$. In this way the
$\Phi$-field is peaked strongly in field space about the unstable
maximum. On incorporating the hard thermal loop 'tadpole' diagrams
of the $\chi$ (and $\Phi$) fields in the $\Phi$ mass term leads to
the effective action for $\Phi$ quasiparticles,
\[
S^{\rm eff}_{\rm syst}[\Phi] = \int d^4x\left\{
{1\over{2}}\partial_{\mu} \phi_{\rm a}\partial^{\mu} \phi_{\rm a} -
{1\over{2}} m_{\phi}^2(T_0) \phi_{\rm a}^2 -
{\lambda\over{4}}(\phi_{\rm a}^2)^2\right\}
\]
where $m_{\phi}^2(T_0)=-\mu^2 (1-T_0^2/T_{\rm c}^2)= M^2>0$. As in
I, we can take an initial factorised density matrix at temperature
$T_0$ of the form ${\hat\rho}[T_0] = {\hat\rho}_{\Phi}[T_0]
{\hat\rho}_{\chi}[T_0]$, where ${\hat\rho}_{\Phi}[T_0]$ is
determined by the quadratic part of $S^{\rm eff}_{\rm syst}[\Phi
]$ and ${\hat\rho}_{\chi}[T_0]$ by $S_{\rm env}[\chi ]$. That is,
the many $\chi_{\rm n}$ fields have a large effect on $\Phi$, but
the $\Phi$-field has negligible effect on the $\chi_{\rm n}$. This
is crucial for the validity of our approximation. More details are
given in I.

In fact, with defects in mind, we need to perform a further
factorisation to isolate the long-wavelength modes of the system
field. As we stressed, the decohering agents are not only the
environment fields $\chi_{\rm n}$ that have been introduced
already but also the short-wavelength modes\cite{fernando2} of the
self-interacting $\Phi$-field.

As in I, we split the fields initially as $\Phi = \Phi_< + \Phi_>$
(and $\phi_{\rm a} = \phi_{<\rm a} + \phi_{>\rm a}$) where the system-field
$\Phi_{<}$ contains the modes with wavelengths longer than the
critical value $\mu^{-1}$, and the bath or environment-field
$\Phi_{>}$ contains wavelengths shorter than $\mu^{-1}$. This is a
natural division, since modes with wavelengths shorter than
$\mu^{-1}$ are stable, modes with longer wavelengths unstable.
Further, the short-wavelengths explore the interior of classical
vortices, irrelevant for counting them. However, once the power in
the fluctuations has moved to long-wavelength modes, with
wavenumber $k_0<\mu$, the exact delineation is unimportant, and
shorter wavelength modes can be included in the 'system'. With
this in mind a more realistic decomposition into 'system' and
'environment' is not (\ref{action1}), but
\[
S[\Phi , \chi ] = S_{\rm syst}[\Phi_<] + S_{\rm env}[\chi,\Phi_> ]
+ S_{\rm int}[\Phi_< ,\chi,\Phi_> ],
\]
with $\mu$ the demarcating momentum, where

\[
S_{\rm syst}[\Phi_<] = \int d^4x\left\{ {1\over{2}}\partial_{\mu}
\phi_{<\rm a}\partial^{\mu} \phi_{<\rm a} + {1\over{2}}\mu^2
\phi_{<\rm a}^2 - {\lambda\over{4}}(\phi_{<\rm a}^2)^2\right\},
\]
and
\begin{eqnarray}
S_{\rm env}[\Phi_>,\chi ] &=&\int d^4x\left\{
{1\over{2}}\partial_{\mu} \phi_{>\rm a}\partial^{\mu} \phi_{>\rm
a} - {1\over{2}}\mu^2 \phi_{>\rm a}^2  - {\lambda\over{4}}
(\phi_{>\rm a}^2)^2\right\}\nonumber
\\
&+& \sum_{\rm n=1}^N\int d^4x\left\{
{1\over{2}}\partial_{\mu}\chi_{\rm n}
\partial^{\mu}
\chi_{\rm n} - {1\over{2}} m^2 \chi^2_{\rm n}\right\}, \nonumber
\\
S_{\rm int}[\Phi_{<} ,\Phi_>,\chi ] &=& - \sum_{\rm
n=1}^N\frac{g}{8\sqrt{N}} \int d^4x \,\phi_{<\rm a}^2 (x)
\chi^2_{\rm n} (x) \nonumber \\
 &-&\frac{\lambda}{2}\int d^4x ~[\phi_{<\rm a}^2(x)
\phi^2_{>\rm b}(x)+2\phi_{<\rm a}(x) \phi_{<\rm b}(x)\phi_{>\rm
a}(x) \phi_{>\rm b}(x)]. \nonumber
\end{eqnarray}
All terms omitted in the expansion are not relevant for the
one-loop calculations for the long-wavelength modes that we shall
now consider.

\section{The simplest transition: Instantaneous quench}

For simplicity we repeat our assumptions in I, in adopting an {\it
instantaneous} temperature quench from $T_0$ to $T_{\rm f}=0$ at
time $t=0$, in which $m_{\phi}^{2}(T)$ changes sign and magnitude
instantly, concluding with the value $m_{\phi}^2=-\mu^2$, $t>0$ (and
beginning with the value $m_{\phi}^2(T_0)=m^2=O(\mu^2)$, $t<0$).  We
stress that $m_{\phi}$ is the renormalised mass, containing the
temperature dependent interactions with all fields.

\subsection{The influence of the environment}

As we observed, the most visible signal that the transition has
occurred will be the presence of topological defects, solutions to
the {\it classical} equations of motion $\delta S[\Phi,\chi ] =
0$, with $\chi_{\rm n} = 0$. These are global vortices in the
field, around which the field phase $\theta$ ($\Phi ({\bf x}) =
{\mbox h}({\bf x}) e^{i\theta ({\bf x})}$) changes by $2\pi$.
Considered as tubes of 'false' vacuum, with cold thickness
$O(\mu^{-1})$, they have energy per unit length $\sigma
=O(\mu^2/\lambda)$ (up to multiplicative logarithmic terms $\ln
(\mu\xi_{\rm def})$, where $\xi_{\rm def}$ is vortex separation).
In particular, the field $\Phi$ vanishes along the vortex core. We
can therefore use line zeroes to track classical defects
\cite{rajantie,findland}. Since classical defects are specific
field profiles we need to be able to distinguish between different
classical system-field configurations evolving after the
transition. As a result, we are only interested in the
field-configuration basis for the density matrix.

Since the effects of (bosonic) environments are cumulative each
contribution to the environment increases the diffusion term and
thereby speeds up the onset of classical behaviour. Thus any part
of the environment sets an {\it upper} bound on $t_D$. For large
(but not infinite) $N$ a one-loop approximation is sufficient for
calculating $t_D$ due to $\chi$-fields alone. It is convenient not
to have complex arguments and we use $\phi$ to denote the real
Cartesian doublet $(\phi_1,\phi_2)$ and $\chi$ to denote
$(\chi_1,\chi_2,..,\chi_{\rm n})$, etc. At time $t>0$ the reduced
density matrix $\rho_{{\rm r}}[\phi^+_<,\phi^-_<, t]=
\langle\phi^+_<\vert {\hat\rho}_r (t)\vert \phi^-_< \rangle$ is
now
\begin{equation}
\rho_{{\rm r}}[\phi^+_<,\phi^-_<, t] = \int{\cal D}\chi ~\int{\cal
D }\phi_> ~ \rho[\phi^+_<,\phi_>,\chi ,;\phi^-_<,\phi_>, \chi ;t],
\end{equation}
where $\rho[\phi^+_<,\phi_>,\chi ;\phi^-_<,\phi_>, \chi ;t]$ is
the full density matrix, ${\cal D}\phi_>={\cal D}\phi_{>1}{\cal
D}\phi_{>2}$ and ${\cal D}\chi =\prod^N_1{\cal D}\chi_{\rm n}$.

The environment will have had the effect of making the system
effectively classical once $\rho_{\rm r}(t)$ is essentially
diagonal. This is very different from the late-time dephasing
effects found in Refs. \cite{salman,mottola}, which rely on
time-averaged diagonalisation. More details are given in paper I
and \cite{lmr}. Quantum interference can then be ignored and we
obtain a classical probability distribution from the diagonal part
of $\rho_{\rm r}(t)$, or equivalently, by means of the reduced
Wigner functional. For weak coupling (see I and \cite{diana})
there will be no 'recoherence' at later times in which the sense
of classical probability will be lost.

The temporal evolution of $\rho_{{\rm r}}[\phi^+_<,\phi^-_<, t]$
is

$$\rho_{{\rm r}}[\phi^+_<,\phi^-_<, t]=
\int {\cal D}\phi_{{\rm i}<}^+ \int {\cal D}\phi_{{\rm i}<}^-
J_{\rm r} [\phi_{{\rm f}<}^+,\phi_{{\rm f}<}^-,t\vert \phi_{{\rm
i}<}^+,\phi_{{\rm i}<}^-,t_0] ~\rho_{\rm r} [\phi_{{\rm i}<}^+
,\phi_{{\rm i}<}^-,t_0],$$
where ${\cal D}\phi^+_< = {\cal D}\phi^+_{<1}{\cal D}\phi^+_{<2}$,
etc. and $J_{\rm r}$ is the reduced evolution operator.

In order to estimate the functional integration
which defines the reduced propagator, we perform a saddle point approximation
\[
J_{\rm r}[\phi^+_{{\rm f}<},\phi^-_{{\rm f}<},t\vert\phi^+_{{\rm
i}<},\phi^-_{{\rm i}<}, t_0] \approx \exp{ i A[\phi^{+\rm cl}_<,
\phi^{-\rm cl}_<]},
\]
where $\phi^{\pm\rm cl}_<$ is the solution of the equation of
motion ${\delta Re
A\over\delta\phi^+}_<\vert_{\phi^+_<=\phi^-_<}=0$ with boundary
conditions $\phi^{\pm\rm cl}_<(t_0)= \phi^{\pm}_{{\rm i}<}$ and
$\phi^{\pm\rm cl}_<(t)=\phi^\pm_{{\rm f}<}$. It is very difficult
to solve this equation analytically. In order to make it tractable
we assume that the system-field contains only one Fourier mode
with $\vec k = \vec k_0$ for the reason indicated earlier, that
the long-wavelength modes, for which $\vert k_0\vert^2 < \mu^2$,
increasingly bunch about a wave-number $k_0 < \mu$  which
diminishes with time. However, unlike in I, we are interested in
more than the $k_0 = 0$ mode.

For such small $k_0$ the classical solution is of the form
\[
\phi^{\rm cl}_{<\rm a}(\vec x, s) =  f_{\rm a}(s,t)\cos(\vec k_0 .
\vec x),
\]
where $f_{\rm a}(s,t)$ satisfies the boundary conditions $f_{\rm
a}(0,t)= \phi_{\rm i<\rm a,}$ and $f_{\rm a}(t,t) = \phi_{\rm f<\rm a}$.
Qualitatively, $f_{\rm a}(s,t)$ grows exponentially with $s$ for $t\leq
t_{\rm sp}$, and oscillates for $ t_{\rm sp}<s<t$ when $t>t_{\rm
sp}$. For $t\leq t_{\rm sp}$ we approximate it by
\begin{equation}
f_{\rm a}(s,t) = \phi_{\rm i<\rm a}{\sinh[\omega_0 (t - s)]
\over{\sinh(\omega_0 t)}} + \phi_{\rm f<\rm a}{\sinh(\omega_0 s)
\over {\sinh(\omega_0 t)}}
\end{equation}
where $\omega_0^2 = \mu^2 - k_0^2$.

We saw in I (see also \cite{Karra}) that the linear approximation
is reasonable until almost $t_{\rm sp}$, where the spinodal time
$t_{\rm sp}$ is defined as the time for which $\langle
|\Phi_<|^2\rangle_t\sim \eta^2$. That is, $t_{sp}$ is the time it
takes for the field to populate the ground states of the model. As
a result, $t_{sp}$ is given by
            \begin{equation}
            \exp [2\mu t_{\rm sp}] \approx  {\cal O}( \frac{\eta^2}{\mu
            T_{\rm c}}).
            \label{exptsp}
            \end{equation}
The exponential factor in Eq.(\ref{exptsp}), as always, arises
from the growth of the unstable long-wavelength modes. The factor
$T_{\rm c}^{-1}$ comes from the $\coth(\beta\omega /2)$ factor
that encodes the initial Boltzmann distribution at temperature
$T_0\gtrsim T_{\rm c}$. Thus,
            \begin{equation}
            \mu t_{\rm sp} \sim
            \ln (\frac{\eta}{\sqrt{\mu T_{\rm c}}}).
             \label{tsp}
            \end{equation}

As in paper I, it is sufficient to calculate the correction to the
usual unitary evolution coming from the noise kernel. For clarity
we drop the suffix ${\rm f}$ on the final state fields. If $\Delta
= (|\Phi^{+}|^2 - |\Phi^{-}|^2)/2$ for the {\it final} field
configurations, then the master equation for $\rho_{\rm r}
(\phi^+_<,\phi^-_<, t)$ is

\begin{equation}
i {\dot \rho}_{\rm r} = \langle \phi^+_<\vert [H,{\hat\rho}_{\rm
r}] \vert \phi^-_<\rangle - i V \Delta^2 D(k_0, t) \rho_{\rm r}+
... \label{master}
            \end{equation}
The time dependent diffusion coefficient $D_{\chi}(k_0,t)$ that
determines the effect of the $\chi$-fields on the onset of
classical behaviour acquires a contribution $D_{\rm n}(k_0, t)$ from
each field $\chi_{\rm n}$,

\begin{equation}
D_{\rm n}(k_0, t)= \frac{g^2}{16N}\int_0^t ds ~ u(s,t) \left[ {\rm
Re}
G_{++}^2(2k_0; t-s)
+ 2 {\rm Re}G_{++}^2(0; t-s)\right].
\label{diff}
\end{equation}
where, for the case in hand of an instantaneous quench, $u(s,t) =
\cosh^2\omega_0(t- s)$ when $t \leq t_{\rm sp}$, and is an
oscillatory function of time when $t>t_{\rm sp}$. The $G_{++}$ are
the long-wavelength correlation functions of the $\chi$-fields for
the appropriate contours in the closed time-path.

The contribution from the explicitly environmental $\chi$-fields,
$D_{\chi}(k_0, t)=\sum_{\rm n} D_{\rm n}(k_0, t)$, takes the form
\begin{equation}
D_{\chi}(k_0, t)\sim \frac{g^2T_0^2}{\mu^3}\omega_0~ \exp
[2\omega_0 t], \label{D(t)}
\end{equation}
largely from the end-point behaviour at $s=0$ of the integral (\ref{diff}).
For $t>t_{\rm sp}$ the diffusion coefficient stops growing, and
oscillates around $D(k_0,t=t_{\rm sp})$.

The environment is also constituted by the short-wavelength modes
 $\Phi_>$ of the self-interacting field. This gives an additional one-loop
contribution to $D(k_0, t)$ with the same $u(s)$ but a $G_{++}$
constructed from the short-wavelength modes of the $\Phi$-field as
it evolves from the top of the potential hill. Omitting this mode
in its calculation just means that  any decoherence time $t_D$
obtained from the $\chi_{\rm n}$ alone will be an upper bound on
the true decoherence time. An estimate of the $\Phi$-field
contribution, based on the inclusion of tadpole diagrams alone,
suggests that, with no $1/N$ damping, the $\Phi_>$ modes have the
same effect on the dissipation, qualitatively, as all the
environmental fields $\chi_{\rm n}$ put together. At an order of
magnitude level we can ignore the $\Phi_>$ modes in the
calculation of $t_D$, for which the $\chi$-fields alone give a
strong bound. We cannot ignore them in the calculation of the
defect density, since it is the coarse-graining of the
order-parameter field that renders the line-zero density finite.

We estimate $t_D$ by solving (in the one-loop approximation) for
the off-diagonal elements of the reduced density matrix
\begin{equation}
 \rho_{\rm r}[\phi^+_{<}, \phi^-_{<}; t] \lesssim
\rho^{\rm u}_{\rm r}[\phi^+_{<}, \phi^-_{<}; t]
~~\exp \bigg[-V\Gamma\int_0^t ds ~D_{\chi}(k_0, s) \bigg],
\label{rho}
\end{equation}
where $\rho^{\rm u}_{\rm r}$ is the solution of the unitary part
of the master equation (i.e. without environment). In (\ref{rho})
$\Gamma = O(\mu^4(\bar\phi\delta )^2)$, $\bar\phi = (|\Phi_{<}^+|
+ |\Phi_{<}^-|)/2\mu$ and $\delta = (|\Phi_{<}^+| -
|\Phi_{<}^-|)/2\mu$. $V$ is understood as the minimal volume
inside which there is no possibility for coherent superpositions of
macroscopically distinguishable states for the field. We take this
as ${\cal O}(\mu^{-3})$  since $\mu^{-1}$ is the thickness of an
individual vortex. Inside this volume we do not discriminate
between field amplitudes which differ by $ {\cal O}(\mu) $, and
therefore, as in I, take $\delta^2 \sim {\cal O}(1)$. Similarly,
we set $\bar\phi^2\sim {\cal O}(\alpha /\lambda)$, where
$\lambda\leq\alpha\leq 1$ is to be determined self-consistently.

The decoherence of the long-wavelength $k_0$-mode by the
environment occurs when the non-diagonal elements of the reduced
density matrix are much smaller than the diagonal ones. In
(\ref{rho}) this corresponds to when
\begin{equation}
1\gtrsim V\Gamma \int_{0}^{t_{D}} ds ~D_{\chi}(k_0, s),
\label{dchi}
\end{equation}
since $\rho^{\rm u}_{\rm r}[\phi^+_{<}, \phi^-_{<}; t]$ is
increasingly independent of $\delta$. See I for more details.
Because the diagonalisation of $\rho_r(t)$ occurs in time as an
{\it exponential} of an {\it exponential}, decoherence occurs
extremely quickly, at time\cite{fernando1}
\begin{equation}
\omega_0 t_D\gtrsim\ln (\frac{\eta}{T_{\rm c}\sqrt{\alpha}}).
\label{tD}
\end{equation}
For $\omega\approx\mu$, the value of $\alpha$ is determined as
$\alpha \simeq \sqrt{\mu/T_{\rm c}}$ from the condition that, at
time $t_D$, $\langle |\phi|^2\rangle_t\sim\alpha\eta^2$. Since
$\alpha\ll 1$, in principle, the field has not diffused far from
the top of the hill before it is behaving classically.

It follows from Eq.(\ref{tD}) that the upper bound on $t_D$ and,
we assume, $t_D$ itself, increases as $k_0\rightarrow \mu$,
although we need a better approximation to see how, in detail.
However, we stress that, as far as counting vortices is concerned,
all that matters is how the power in the field fluctuations is
distributed. The distance between defects is the relevant
wavelength, and not defect size. With $\omega_0\approx \mu$ for
the relevant  $k_0^2 = O(\mu/t_D)$ it follows that $1 < \mu t_D
\leq \mu t_{\rm sp}$, with
\begin{equation}
\mu t_{\rm sp} -\mu t_D\simeq \frac{1}{4}\ln (\frac{T_{\rm
c}}{\mu}). \label{dt}
\end{equation}

Whereas the environment is very effective at decohering adjacent
field configurations, it has much less impact on the diagonal
matrix elements
\[
P_t[\phi_<] = \langle \phi_<|\rho_r (t)|\phi_<\rangle ,
\]
which give the relative probability that the field takes the value
$\phi_<$ at time $t$, once the theory is classical. This is all
that is needed to calculate equal-time correlation functions of
$\phi_<$, and thereby\cite{halperin,maz} to calculate the density
of classical defects by itemising their zero-field cores.

We know that, for very early times, the Gaussian approximation for
$\rho_r[\phi_<,\phi_<,t]$ is valid, although we should not
interpret it as a probability then. We shall now argue that can we
use the approximation $P_t[\phi_<]\simeq \rho^{\rm
u}_r[\phi_<,\phi_<,t]$, compatible with (\ref{rho}) at least until
time $t_D$. That is, although the environment is crucial in making
$\rho_t$ off-diagonal, it has much less effect on the diagonal
matrix elements, which are the ones we use for calculation.

\section{Classical Equations}

Our analysis of the onset of decoherence shows that it makes
little sense to talk about line zeroes of the field before $t_D$
of Eq.(\ref{tD}), since they would be expected to suffer from
quantum interference, as well as having a density that is strongly
dependent on the scale separating $\phi_<$ from $\phi_>$.

By the time $t_{\rm sp}$ the Gaussian approximation has broken
down, and the Goldstone field phase $\theta$  will have decoupled
from the heavy h-mode. From here onwards it is the massless
Goldstone modes whose causal propagation controls field ordering.
 This is the last ingredient in turning a line-zero into a
classical vortex.

This requires both classical probabilities and classical
equations. Let us consider probabilities first. The reduced Wigner
functional is defined as
\begin{equation}
W_t[\phi_{<},\pi_{<}] = \int{\cal D} \eta_{<}
~~e^{i\pi_{<}\eta_{<}}~
~ \langle\phi_{<}- \eta_{<}|\rho_{\rm r}(t)| \phi_{<}+
\eta_{<}\rangle
\end{equation}
whereby
\begin{equation}
P[\phi_{<}]_t = \int {\cal D}\pi_{<}\,W_t[\phi_{<},\pi_{<}]
=\langle\phi_{<}|\rho_{\rm r}(t)|\phi_{<}\rangle
\label{p}\end{equation} is the probability density for field
configurations.

For $t\geq t_D$ of (\ref{tD}), $W_t[\phi_{<},\pi_{<}]$ is
positive, at least for long-wavelengths, and can be identified
with the Fokker-Planck probability distribution function $P^{\rm
FP}_t[\phi_{<},\pi_{<}]$, from which $P[\phi_{<}]_t$ can be
equally identified as the Fokker-Planck probability
\begin{equation}
P^{\rm FP}[\phi_{<}]_t = \int {\cal D}\pi_{<}\,P^{\rm FP}_t
[\phi_{<},\pi_{<}].
\end{equation}

Suppose we can calculate $P[\phi_{<}]_t$ for times $t\geq t_D$.
This permits us to calculate the equal time n-point correlation
functions
\begin{equation}
G_{<ab..c}^{(\rm n)}({\bf x_1,..,x_{\rm n},t})= \int{\cal
D}\phi_{<} P^{\rm FP}_t[\phi_{<}]\phi_{<\rm a} ({\bf
x_1})...\phi_{<\rm c}({\bf x_{\rm n}}).\label{n}
\end{equation}
As we know\cite{halperin,maz}, equal-time correlators are all we
need to calculate densities of zeroes (line-zeroes, etc.).
Moreover, the dominance of $P[\phi_{<}]_t$ by long-wavelength
modes permits the adoption of a single decoherence time $t_D$ from
Eq.(\ref{dt}). However, if all we are going to use is
$P[\phi_{<}]_t$, the diagonal matrix element of $\rho_t$, there is
no real need to construct the Wigner functional. We can just do a
calculation of $P[\phi_{<}]_t$ from the start, along the lines of
Ref.\cite{boya}.

To see the appearance of individual defects is more difficult, and
requires classical stochastic (Langevin) equations. We have yet to
complete the analysis. However, as a starting point let us pretend
that the only environment is the $\chi$ fields. In the usual
fashion, one can regard the imaginary part of $\delta A$ as coming
from a single noise source $\xi (x)$, with a Gaussian functional
probability distribution given by \cite{fernando2,GR,GM}
\begin{equation}
{\cal P}[\xi ]= N_{\xi}\exp\bigg\{-{1\over{2}}\int d^4x\int d^4y
~\xi\Big[ g^2 N\Big]^{-1}\xi\bigg\},
\end{equation}
where $N_{\xi}$ is a normalization factor, and $N(x-y)\propto Re
\,G_{++}^2(x-y)$ for the single $\chi$-loop. Indeed, we can write
the imaginary part of the influence action as a functional
integral over the Gaussian field $\xi (x)$,

\begin{eqnarray}
&&\int  {\cal D}\xi {\cal P}[\xi ]\exp{\left[\int d^4x -i
\bigg\{\Delta (x) \xi (x) \bigg\}\right]}\nonumber
\\ &=& \exp{\bigg\{-i\int d^4x\int d^4y \ \Big[\Delta (x)
~g^2 N(x,y)~ \Delta (y)\Big] \bigg\}}.\nonumber
\end{eqnarray}

Therefore, the coarse-grained effective action (see I for
definitions) can be rewritten as
\begin{equation}A[\phi^+,\phi^-]=-{1\over{i}} \ln
\int {\cal D} \xi
 P[\xi]
\exp\bigg\{i S_{\rm eff}[\phi^+,\phi^-, \xi]\bigg\},
\end{equation}
where
\begin{equation}
S_{\rm eff}[\phi^+,\phi^-,\xi ]= {\rm Re} A[\phi^+,\phi^-]- \int
d^4x\Big[\Delta (x) \xi (x) \Big].
\end{equation}
Taking the functional variation
\begin{equation}
\left.{\delta S_{\rm eff}[\phi^+,\phi^-, \xi ] \over{\delta
\phi_{\rm a}^+}}\right\vert_{\phi_{\rm a}^+=\phi_{\rm a}^-}=0,
\end{equation}
gives the ``semiclassical Langevin equation'' for the system-field
\cite{fernando2,GR,GM} (up to factors $O(1)$)
\begin{equation}
\Box \phi_{\rm a} (x) - \tilde{\mu}^2 \phi_{\rm a} +
\tilde{\lambda} \phi_{\rm a}(x)\phi_{\rm b}^2(x)
+ g^2 \phi_{\rm a} (x) \int d^4y ~ K(x,y)~ \phi_{\rm b}^2(y) =
\phi_{\rm a} (x)\xi (x), \label{lange2}
\end{equation}
where we have assumed only quadratic interactions, and $K$ is the
(assumed) common mass retarded loop, arising from the real part of
$\delta A$. $\tilde{\mu}$ and $\tilde{\lambda}$ are the 'renormalised'
constants by virtue of the coupling with the environment.

To estimate the size of $g^2 \phi_{\rm a} (x) \int d^4y ~ K(x,y)~
\phi_{\rm b}^2(y)$ in comparison with the $\mu^2 \phi_{\rm a}$
term, we note first that
\begin{equation}
K\propto T^2\mu^2\sim T^2_{\rm c}\mu^2\sim \mu^4/\sqrt{N}\lambda
\end{equation}
for high $T$.
Further, with relevant distance scales no larger
than $\mu^{-1}$ then, crudely, at time $t_{\rm sp}$,
\begin{eqnarray}
&&|g^2\int d^4y ~ K(x,y)~ \phi_{\rm b}^2(y)|\nonumber
\\
&&\leq g^2\int d^4y ~ |K(x,y)|~ \phi_{\rm b}^2(t_{\rm sp}) \nonumber
\\
&& \sim \lambda^2 \frac{T_{\rm c}^2}{\mu} t_{\rm
sp}\phi_{\rm b}^2(t_{\rm sp})\sim
 \mu^2\frac{\mu t_{\rm sp}}{N^{1/2}}\ll \mu^2,
\end{eqnarray}
for large enough $N$.  Since the bound is even smaller at earlier
times this suggests that the dissipative term is negligible, in
comparison to the mass term, until time $t_{\rm sp}$. Equivalently,
for weak coupling the damping due to the thermal environment has a
negligible effect on the quasi-particle mass at these early times.

Even though $\langle\phi_{\rm a} (x)\xi (x)\rangle\neq 0$,
\begin{eqnarray}
\langle\phi_{\rm a} (x)\xi (x)\rangle &\leq &
\sqrt{\langle\phi_{\rm a}(x)\phi_{\rm a}
(x)\rangle \langle\xi (x)\xi (x)\rangle} \nonumber \\
&\leq &\eta \sqrt{\langle\xi (x)\xi (x)\rangle}
\end{eqnarray}
(with no summation over ${\rm a}$) which, from the behaviour of the
noise kernel N, can be bounded as
\[
\langle\phi_{\rm a} (x)\xi (x)\rangle_{\xi}\leq \mu^2\phi_{\rm
a} (x).
\]
It follows that ${\phi_{\rm a}}$ satisfies the classical equation
\begin{equation}
\Box \phi_{\rm a}(x) - \tilde{\mu}^2 \phi_{\rm a}(x) +
\tilde{\lambda} \phi_{\rm a}(x)\phi_{\rm b}^2(x) = 0,
\label{classical}
\end{equation}
to a good approximation for times $t\lesssim t_{\rm sp}$. For such
early times the non-linear term in (\ref{classical}) can be
neglected. For times later than $t_{\rm sp}$ neither perturbation
theory nor more general non-Gaussian methods are valid. For
example, in quantum mechanical models (without environment) {\it
seventh} order calculations in a self-consistent
$\delta$-expansion (that effectively plays the role of a $1/N$
expansion in large-N calculations \cite{boya}) ceases to work once
the symmetry has been fully broken\cite{hugh}. For such times we
need to solve for classical vortices, rather as we would solve for
the evolution of vortex tangles in superfluid $^4 He$.

In terms of the radial and angular fields Eq.(\ref{classical})
becomes
\begin{eqnarray}
\Box {\mbox h} +[-\tilde{\mu}^2 + \tilde{\lambda}{\mbox h}^2 -
(\partial_{\mu}\theta\partial^{\mu}\theta)]{\mbox h} &=& 0\nonumber \\
\partial_{\mu}({\mbox h}^2\partial^{\mu}\theta) &=& 0.
\label{classical2}
\end{eqnarray}

At early times we have seen that $\phi_1$ and $\phi_2$ are
independent, and there are no Goldstone modes. However, by the
time $t\approx t_{\rm sp}$, when $h^2\sim\eta^2$ and is slowly
varying, it follows from the second of the equations that the
Goldstone modes have appeared.  More generally, it is through the
coupling of the Goldstone and Higgs ($h$) modes that classical
defects appear as solutions to (\ref{classical}), in a way that
was denied at early times.

We stress that, in deriving Eq.(\ref{classical}), we have not had
to restrict ourselves to any particular modes in k-space. However,
although Eq.(\ref{classical}) looks to be valid at all times, it
is only a sensible equation once field configurations have
well-defined probabilities associated to them {\it i.e. after}
decoherence. Here lies a difficulty in that we have seen that
$t_D$ depends on wavelength, with shorter wavelengths taking
longer to become classical. Although the long-wavelengths with
most power in their fluctuations, that determine the separation of
vortices, have become classical by time $t_{\rm sp}$, this is an
ensemble statement that cannot be applied easily to individual
vortex solutions to (\ref{classical2}). Further,
Eq.(\ref{classical}) couples $\phi_<$ to $\phi_>$. The effect of
including the short-wavelength modes $\phi_>$ in the environment
is, most likely, only a qualitative change at a consistent
one-loop level, by augmenting $K$ in (\ref{lange2}) with a
comparable term $K_{\phi}$. There will be a further comparable
non-diagonal term $\lambda^2\phi_{\rm b}\int K_{\phi}\,\phi_{\rm a}
\phi_{\rm b}$. We
anticipate that the Langevin equations for the stochastic $\phi_<$
will be just as one would guess from a mode decomposition of
(\ref{classical2}) with short-wavelength modes discarded, and
terms with $K$ and $N$ equally ignorable.  This will be pursued
elsewhere.

\section{Classical defects}

\subsection{Line densities and the Gaussian approximation}

As we observed, the most visible signal that the transition has
occurred will be the presence of classical topological vortices
(strings), radial solutions to the equations of motion $\delta
S[\phi,\chi ] = 0$, with $\chi_{\rm n} = 0$. These are line
defects in the field, around which the field phase $\theta$ ($\phi
= {\mbox h} e^{i\theta}$) changes by $2\pi$.  We can therefore use
line zeroes to track classical
vortices\cite{rajantie,findland,al}. [If we had broken an $O(3)$
symmetry ($a=1,2,3$) the corresponding defects would be global
monopoles, giving qualitatively similar conclusions.]

The {\it total line-zero density} $\bar{z}({\bf x})$ for the
long-wavelength mode fields is\cite{halperin}
\begin{equation}
\bar{z_{i}}({\bf x}) = \delta^{2}[\phi_< ({\bf
x})]|\epsilon_{ijk}\partial_{j} \phi_{<1}({\bf x})
\partial_{k}\phi_{<2}({\bf x})|. \label{rhobar}
\end{equation}
With line zeroes as the intersections of planes of zeroes of
$\phi_1$ and $\phi_2$ separately, we have a phase change of $\pm
2\pi$ around the lines of intersection, but for exceptional
circumstances.

>From Eqs. (\ref{p}) to (\ref{n}) the evolution of the two-field
correlator ($G_<^{(2)}({\bf x},0,t) = G_<(r,t)$) can be written in
terms of the unitary (reduced) density matrix $\rho_{\rm r}^{\rm
u}$. Thus,
\[
G_<(r, t)=\langle\phi_< ({\bf x})\phi_<^*({\bf 0})\rangle_t
=\frac{1}{2}\delta_{ab}\langle\phi_{<\rm a} ({\bf x})\phi_{<\rm b}({\bf
0})\rangle_t
\]
is well understood. The unstable long-wavelength modes grow
exponentially fast. If the power spectrum of the field
fluctuations is defined by
\begin{equation}
G_<(r, t)= \int_{k<\mu} {d^3k\over{(2\pi )^3}}\,P(k,t)\, e^{i{\bf k}.{\bf
x}}
\end{equation}
then $k^2P(k,t)$ rapidly develops a 'Bragg' peak at $k^2 =
k_0^2={\cal O}(\mu/t)$.

While the Gaussian approximation\cite{halperin,maz} is satisfied,
the line-zero ensemble density $n_{\rm zero}(t)$ is determined
completely by the {\it short-distance} behaviour of $G_<(r, t)$ as
\begin{equation}
n_{\rm zero}(t) = \; \langle\bar{z_{i}}({\bf x})\rangle_{t} =
\frac{-1}{2\pi}\frac{G''_<(0, t)}{G_<(0, t )}. \label{ndeff}
\end{equation}
Since $G_< (r,t)$ has short-wavelength modes removed, $G_<(0,t)$
is finite.

Further, for this period when the self-consistent approximation is
valid, the field energy $\langle E\rangle_t$ of the system field
$\phi_<$ in a box of volume ${\cal V}$ becomes
\begin{eqnarray}
\langle E\rangle_t &=& {\cal V}[\langle |\nabla\phi_<|^2\rangle_t
+\lambda(\eta^2 - \langle |\phi_<|^2\rangle_t)^2]\nonumber
\\ &=&
{\cal V}[2\pi n_{\rm zero}(t) G_<(0,t)+ \lambda(\eta^2 - G_<(0,t))^2]
\nonumber \\
 &=& 2\pi L_{\rm zero}(t)G_<(0,t) +{\cal V}[\lambda(\eta^2
- G_<(0,t))^2]. \label{E3}
\end{eqnarray}
As before, $\chi$-field fluctuations are absorbed in the
definition of $\mu^2$. Eq.(\ref{E3}) is obtained by using
(\ref{ndeff}), and $L_{\rm zero}(t)={\cal V} n_{\rm zero}$ is the
total length of line zeroes, on a scale $\mu^{-1}$, in the box of
volume ${\cal V}$.

We understand Eq.(\ref{E3}) as follows. Suppose it were valid from
time $t=0$, when $G_<(0,0) = O(\mu^2)$, until time $t=t_{\rm sp}$. At
early times most of the system field energy (proportional to
${\cal V}$) is in fluctuations not associated with line zeroes,
arising from the field potential. As time passes their energy
density decreases as the system field approaches its
post-transition value, becoming approximately zero. In part, this
is compensated by the term, arising from the field gradients,
proportional to the length $L_{\rm zero}$ of line zeroes, whose energy
per unit length increases from $O(\mu^2)$ to $O(\eta^2)$. At time
$t_{\rm sp}$, when the fluctuation energy can be ignored, we
find\cite{RKK}
\begin{equation}
\langle E\rangle_t \sim L_{\rm zero}\sigma, \label{E}
\end{equation}
essentially the energy required to produce a vortex tangle of
length $L_{\rm zero}$ (up to $O(1)$ factors from the logarithmic
tails). Although these line zeroes have the topological charge and
energy of vortices, they are not yet fully-fledged defects.
Initially, $G_<(r,t)$ is very dependent on the value of the
cut-off. As a result, line-zeroes are extremely fractal, with a
separation proportional to the scale at which they are viewed, and
are certainly not candidates for defects. Once the 'Bragg peak' at
$k=k_0$ is firmly in the interval $k<\mu$, $n_{\rm zero}(t)$ becomes
insensitive to a cut-off $O(\mu^{-1})$. This means that
line-zeroes are straight at this scale, although they can be
approximately random walks at much larger scales. For sufficiently
weak couplings their density this has happened  by time $t_{\rm sp}$.
The final coupling of radial to angular modes that turns these
proto-vortices into vortices incurs no significant energy change,
and $n_{\rm zero}$ of (\ref{ndeff}) is a reliable guide for the
initial vortex density.

\section{Final comments}

The mechanism for vortex production that we have proposed here has
two parts. Firstly, the environment renders the order-parameter
field classical at early times $t_D \lesssim t_{\rm sp}$, by or before
the transition is complete. Secondly, classical defects evolve
from line-zeroes whose resultant density can already be inferred
in the linear regime, but whose specific attributes are a
consequence of the non-linear Langevin equations  at the spinodal
time $t_{\rm sp}$.

This  is very different from traditional explanations, and we
conclude with a brief summary of them.

An early explanation, originally due to Kibble\cite{kibble1}, and
still of common currency, is that thermal fluctuations in the
Ginzburg regime might lead to the production of vortices, again at
early times. The reason is the following: once we are below
$T_{\rm c}$, the Ginzburg temperature $T_{\rm G}<T_{\rm c}$
signals the temperature above which there is a significant
probability for thermal fluctuations between one degenerate
groundstate and another on the scale of the correlation length at
that temperature. That is, the thermal energy in such a
fluctuation matches the energy required to pass over the hump of
the unstable minimum. This picture presupposes a slow quench, and
cannot be accommodated in the instantaneous quench approximation
that we have used here. However, our suggestion that defects only
appear at, or about, the spinodal time $t_{\rm sp}$ at a density
given by the density of line zeroes is totally at variance with
this picture, and thermal activation is not the relevant
mechanism.

In fact, this was recognised early by Kibble himself\cite{kibble2}, with
his later emphasis on strong causal bounds. It had been noted
earlier\cite{kibble1} that the field must behave independently in initially
space-like separated regions. When these domains with different vacua become
causally connected we expect defects to link them. This gives
late-time predictions that we cannot address with our early-time
analysis. A more powerful variant on this theme identifies the
time at which defects first appear as the time the adiabatic
(long-distance) correlation decreases at the speed of light. This
time, and the resultant density, depends on the quench time
$\tau_{\rm Q}$. There are also difficulties with this in that we have
seen that the separation of line-zeroes are obtained from the
short-distance behaviour of $G_<(r,t)$. The argument was not posed
for instantaneous quenches of the type discussed above, which
correspond to $\tau_{\rm Q} = O(\mu^{-1})$ for which, if taken
literally, it would give defect formation from time $O(\mu^{-1})$,
again incorrect.  We are in the process of extending the analysis
above to slower quenches to make the comparison with the
predictions of \cite{kibble2} more useful.

Subsequently, two approaches have been adopted. In the first,
motivated by condensed matter physics, for which similar causal
arguments apply\cite{zurek1}, phenomenological classical
stochastic Langevin equations have been proposed (e.g. see
\cite{zurek2}) for the evolution of the fields from early times.
These are the counterpart of the phenomenological time-dependent
Ginzburg-Landau equations of condensed matter and assume classical
probabilities from the beginning. However, unlike the intermediate-time
Langevin equations derived previously, these equations do not have
multiplicative dissipation and multiplicative
noise\cite{fernando2,GR,GM}.
The second approach is intrinsically quantum mechanical, treating
the quantum field as a closed system.  Although classical
correlations are present in the localisation of the Wigner
functional\cite{guthpi}, even for such closed systems, it is
difficult to identify defects easily, given that there are no
classical probabilities at early times. At best, there is
late-time dephasing\cite{salman}. The mechanism proposed in this
paper is totally different.

\acknowledgments F.C.L. and F.D.M. were supported
by Universidad de Buenos Aires, CONICET (Argentina), Fundaci\'on
Antorchas and ANPCyT. R.J.R. was supported in part by the COSLAB
programme of the European Science Foundation. We also thank the
organisers of the Peyresq meeting.


\begin{references}


\bibitem{kibble1} T.W.B. Kibble, {\it J. Phys.}\, {\bf A9}, 1387
(1976).


\bibitem{zurek1} W.H. Zurek, {\it Physics Reports}\, {\bf 276}, 177, (1996).


\bibitem{kibble2} T.W.B. Kibble, {\it Physics Reports}\, {\bf 67}, 183 (1980).


\bibitem{rajantie} A. Rajantie, ``{\it Formation of topological defects in
gauge field theory}'', hep-ph/0108159.


\bibitem{fernando2} F.C. Lombardo and F.D. Mazzitelli, Phys. Rev. {\bf D53},
2001 (1996)


\bibitem{diana} F.C. Lombardo, F.D. Mazzitelli, and D. Monteoliva,  Phys.
Rev. {\bf D62}, 045016 (2000).


\bibitem{fernando1}F.C. Lombardo, F.D. Mazzitelli and R.J. Rivers,
 these proceedings.


\bibitem{lmr}F.C. Lombardo, F.D. Mazzitelli and R.J. Rivers,
Phys. Lett. {\bf B523}, 317 (2001).



\bibitem{findland} R.J. Rivers, ``{\it Zurek-Kibble Causality Bounds in
Time-Dependent Ginzburg-Landau Theory and Quantum Field Theory}'',
Journal of Low Temperature Physics, June/July 2001 as a general
article in the proceedings of the ULTI conference, Finland, 2001.
cond-mat/0105171



\bibitem{salman} S. Habib, Y. Kluger, E. Mottola, and J.P. Paz,
Phys. Rev. Lett. {\bf 76}, 4660 (1996)


\bibitem{mottola}  F. Cooper, S. Habib, Y. Kluger, and E.
Mottola, Phys. Rev. {\bf D55}, 6471 (1997)



\bibitem{Karra}G. Karra and R.J. Rivers, Phys. Lett. {\bf B414}, 28 (1997).


\bibitem{halperin} B.I. Halperin, published in {\it Physics of
Defects}, proceedings of Les Houches, Session XXXV 1980 NATO ASI,
editors Balian, Kl\'{e}man and Poirier (North-Holland Press, 1981)
p.816.


\bibitem{maz} F. Liu and G.F. Mazenko, {\it Phys. Rev.}\, {\bf
B46}, 5963 (1992).


\bibitem{boya}  D. Boyanovsky, H.J. de Vega, and R. Holman, Phys. Rev.
{\bf D49}, 2769 (1994); D. Boyanovsky, H.J. de Vega, R. Holman,
D.-S. Lee, and A. Singh, Phys. Rev. {\bf D51}, 4419 (1995); D.
Boyanovsky, D. Cormier, H.J. de Vega, R. Holman, and S. Prem
Kumar, Phys. Rev. {\bf D57}, 2166 (1998)


\bibitem{GR} M. Gleiser and R.O. Ramos, {\it Phys.
Rev.}\,{\bf D50}, 2441 (1994)


\bibitem{GM} C. Greiner and B. Muller, {\it Phys.
Rev.}\,{\bf D55}, 1026 (1997)


\bibitem{hugh} H.F. Jones, P. Parkin,  and D. Winder, Phys. Rev. {\bf D63},
125013 (2001)


\bibitem{al}A.J. Gill and R.J. Rivers, Phys. Rev. {\bf D51}, 6949 (1995)


\bibitem{RKK} E. Kavoussanaki,
R.J. Rivers and G. Karra, {\it Condensed Matter Physics} {\bf 3},
133 (2000).


\bibitem{zurek2} N.D. Antunes, L.M.A. Bettencourt, and W.H. Zurek, Phys. Rev.
Lett. {\bf 82}, 2824 (1999)


\bibitem{guthpi} A. Guth and S.Y. Pi, Phys. Rev. {\bf D32}, 1899 (1991)


\end{references}
\end{document}